\documentclass[]{spie}

\def\simlt{\lower.5ex\hbox{$\; \buildrel < \over \sim \;$}}
\def\simgt{\lower.5ex\hbox{$\; \buildrel > \over \sim \;$}}
\newcommand{\citep}[1]{{\cite{#1}}}

\newcommand{\degree}{\ifmmode {^{\circ}} \else {$^{\circ}$} \fi}
\newcommand{\degrees}{\ifmmode {^{\circ}} \else {$^{\circ}$} \fi}

\newcommand{\unit}[1]{\ifmmode {\rm\ #1\,} \else {$\rm #1$} \fi}
\newcommand{\quarter}{\ifmmode {\frac{1}{4}} \else {$\frac{1}{4}$} \fi}
\newcommand{\arcmin}{\ifmmode {^\prime} \else {$^\prime$} \fi}
\newcommand{\arcsec}{\ifmmode {^{\prime\prime}} \else {$^{\prime\prime}$} \fi}

\newcommand{\lt}{\unit{<}}
\newcommand{\gt}{\unit{>}}

\newcommand{\tten}[1]{\ifmmode {\times 10^{#1}} \else {$\times 10^{#1}$} \fi}
\newcommand{\tentothe}[1]{\ifmmode {10^{#1}} \else {$10^{#1}$} \fi}

\newcommand{\dm}{\unit{pc\ cm^{-3}}}

\newcommand{\doublet}{\ifmmode {\lambda\lambda} \else {$\lambda\lambda$} \fi}
\newcommand{\singlet}{\ifmmode {\lambda} \else {$\lambda$} \fi}

\usepackage{pslatex} 
\usepackage{graphicx} 
\usepackage{epsfig}
\usepackage{hyperref} 
\usepackage{epsf} 
\usepackage{ifpdf}
\usepackage{url}

\newcommand{\ifnotspie}[1]{}

\title{Status of the UC-Berkeley SETI Efforts}

\author{E.~J.~Korpela\supit{a}, D.~P.~ Anderson\supit{a}, R.~Bankay\supit{a}, J.~Cobb\supit{a}, A.~Howard\supit{a},\\
M.~Lebofsky\supit{a}, A.~P.~V.~Siemion\supit{a}, J.~von Korff\supit{b}, and D.~Werthimer\supit{a} 
\skiplinehalf
\supit{a}University of California, Berkeley, CA, 94720 USA;\\
\supit{b}Kansas State University, Manhattan, KS, 66502 USA}

\authorinfo{{\vspace{-0.3in}\\\ \\Email: {\em korpela@ssl.berkeley.edu}\\ Telephone:
+1 510 643-6538\\ URL: \href{http://setiathome.berkeley.edu/}{\em
http://setiathome.berkeley.edu/}}}

\begin{document}

\maketitle

\ifnotspie{
\begin{center}
{\footnotesize
Copyright 2011 Society of
Photo-Optical Instrumentation Engineers. One print or electronic copy may be made for
personal use only.  Systematic reproduction and distribution for any material in
this paper for a fee or for commercial purposes, or modification of the content
of the paper are prohibited.\\\ \\Korpela et al., Proc. SPIE {\bf 8152} (2011)}
\end{center}
}

\begin{abstract} 
We summarize radio and optical SETI programs based at
the University of California, Berkeley.

The SEVENDIP optical pulse search looks for ns time scale pulses at
visible wavelengths.  It utilizes an automated 30 inch telescope, three
ultra fast photo multiplier tubes and a coincidence detector.  The target
list includes F, G, K and M stars, globular cluster and galaxies.

The ongoing SERENDIP V.v sky survey searches for radio signals at the
300 meter Arecibo Observatory.  The currently installed configuration
supports 128 million channels over a 200 MHz bandwidth with $\sim$1.6
Hz spectral resolution.  Frequency stepping allows the spectrometer to
cover the full 300MHz band of the Arecibo L-band receivers.  The final
configuration will allow data from all 14 receivers in the Arecibo L-band
Focal Array to be monitored simultaneously with over 1.8 billion channels.

SETI@home uses the desktop computers of volunteers to analyze over
160 TB of data at taken at Arecibo.  Over 6 million volunteers have run
SETI@home during its 10 year history.  The SETI@home sky survey is 10
times more sensitive than SERENDIP V.v but it covers only a 2.5 MHz band,
centered on 1420 MHz.  SETI@home searches a much wider parameter space,
including 14 octaves of signal bandwidth and 15 octaves of pulse period
with Doppler drift corrections from -100 Hz/s to +100 Hz/s.  SETI@home is
being expanded to analyze data collected during observations of {\em Kepler}
objects of interest in May 2011.

The Astropulse project is the first SETI search for $\mu$s time scale
pulses in the radio spectrum.  Because short pulses are dispersed
by the interstellar medium, and the amount of dispersion is unknown,
Astropulse must search through 30,000 possible dispersions.  Substantial
computing power is required to conduct this search, so the project uses
volunteers and their personal computers to carry out the computation
(using distributed computing similar to SETI@home).
\end{abstract}


\section*{INTRODUCTION}

At the University of California, Berkeley, we have been conducting five
SETI searches that are roughly orthogonal to each other in search space.
These five searches are summarized in Table 1.

\begin{table}[t] 
\begin{center} 
\begin{tabular}{lcll}
\multicolumn{1}{c}{Program Name} & \multicolumn{1}{c}{Timescale} &
\multicolumn{1}{c}{Wavelength} & 
\multicolumn{1}{c}{$ \Delta \nu / \nu $} \\
\hline \hline 
SERENDIP V.v & 1 s & radio & 0.21 \\
SETI@home & 1 ms -- 10 s & radio & 0.0018 \\ 
Fly's Eye & ms & radio & 0.15 \\
Astropulse & $\mu$s & radio & 0.0018 \\
SEVENDIP & ns & optical & 0.8 \\
\hline 
\end{tabular} 
\\\ \\ 
\caption{SETI programs at the University of California, Berkeley}
\end{center}
\end{table}

The SERENDIP V.v sky survey covers a relatively broad range of radio
frequencies, but not as sensitively as SETI@home.  The SETI@home sky
survey is more sensitive and examines a much wider variety of signal
types than SERENDIP, but only covers a narrow band centered on the 21
cm Hydrogen line (a ``magic frequency").  The Astropulse program is the
first search for $\mu$s time scale radio pulses.  The SEVENDIP optical
pulse search is sensitive to low duty cycle ultra-short pulses (eg:
pulsed lasers).

We describe each of these programs below.

\section*{OPTICAL SETI}

There is no clear wavelength choice for SETI.  Microwave, IR and visible
wavelengths all have advantages and disadvantages, depending on what
factors another civilization might choose to optimize (power, size,
bandwidth, and/or beam size).  Although optical photons require more
energy to generate than radio photons, optical beam sizes are typically
much smaller, and directed interstellar communication links can be more
efficient.\cite{Lampton00,Townes61,Townes97}

\section*{SEVENDIP}

The SEVENDIP 
(Search for Extraterrestrial Visible Emissions
from Nearby Developed Populations)
program at Berkeley searches for nanosecond time scale
pulses, perhaps transmitted by a powerful pulsed laser operated by a
distant civilization.  The target list includes mostly nearby F, G, K
and M stars, plus a few globular clusters and galaxies.  The pulse search
utilizes Berkeley's 0.8 meter automated telescope at Leuschner observatory
and specialized instrumentation to detect short pulses.  A similar
instrument has been developed at Harvard University.\cite{Howard99}

The SEVENDIP instrument uses beam splitters to feed light from the
telescope onto three high speed photomultiplier tubes.\cite{Wright01}
These tubes have a rise time of 0.7 ns and are sensitive to 300 - 700
nm wavelengths.  The three signals are fed to high speed amplifiers, fast
discriminators, and a coincidence detector.  Three detectors are needed
to reject ``false alarms," which can be caused by radioactive decay and
scintillation in the PMT glass, cosmic rays, and ion feedback. These
false alarms can happen often in a single PMT, but almost never occur
in three PMT's simultaneously.

The Leuschner pulse search has examined several thousand stars so far,
each star for one minute or more.  The experiment's sensitivity is 1.5
$\times 10^{-17}$~W/m$^2$ for a 1 ns pulse, which corresponds to 1.5
$\times 10^{-28}$~W/m$^2$ average power if the pulse duty cycle is one
nanosecond every 100 seconds.

\section*{THE SERENDIP V.v ARECIBO SKY SURVEY}

The SERENDIP SETI program began 25 years ago; it has gone through four
generations of instrumentation and has observed on 14 radio telescopes.
During these twenty five years, SERENDIP's sensitivity has improved by
a factor of ten thousand and the number of channels has increased from
one hundred to more than one hundred million.\cite{Werthimer97,Bowyer97}

The latest SERENDIP sky survey, SERENDIP V.v, began in earnest in 2009.
Observations are ongoing.  The survey utilizes the National Astronomy and
Ionospheric Center's 305 meter radio telescope in Arecibo, Puerto Rico.
The survey thoroughly covers 25\% of the sky (declinations from -2\degree to +38\degree).
Each of the 10 million beams will be observed an average of three times
during the five year survey.  Multiple observations are needed because
sources may scintillate\cite{Cordes97} or have short duty cycles, and
many of our robust detection algorithms require multiple detections.

The sky survey utilizes real time 128 million channel FFT spectrum
analyzers to search for narrow band radio signals in a 300 MHz band
centered at the 21 cm Hydrogen line (1420 MHz).  The currently installed
system consists of one such instrument.  We are working towards a final
configuration consisting of 14 of these instruments, which will allow
simultaneous analysis of data from all of the 14 receivers of the Arecibo
L-band Focal Array (ALFA). The system has a 0.6 second integration time,
1.6 Hz channel width, and a sensitivity of 10$^{-24}$~W/m$^2$.

SERENDIP V.v conducts observations continuously whenever the ALFA
array is in use for other astronomical observations.  SERENDIP data
analysis is described by Cobb\cite{Cobb00}.  Information on signals
whose power exceeds 16 times the mean noise power are logged along with
baseline power, telescope coordinates, time and frequency.  This data is
transmitted to Berkeley in real time; then, radio frequency interference
(RFI) rejection algorithms are applied to the data, off-line, at UC
Berkeley.  After the RFI is rejected, computers search for candidate
signals.  SERENDIP's candidate detection algorithms are sensitive to
several types of signals, which, individually or combined, may trigger
an event to be noted for further study. These algorithms test for beam
pattern matching, linear drift rates, regularly spaced pulses, multiple
frequencies (particularly those periodic in frequency), and coincidence
with nearby stars, globular clusters, or extra-solar planetary systems.
Every few months, the entire data base is scanned for multiple detections
-- ``signals" that are detected again when the telescope revisits the
same sky coordinates. We test how well these multiple detections fit a
barycentric reference frame.  We also apply another test that allows
much higher frequency separation, which is necessary if transmitters
are not corrected for their planet's rotation and revolution.  Data are
simultaneously sent to Cornell University for analysis using other
techniques.

Potential candidates are scored and ranked by the probability of noise
causing that particular detection. In cases where multiple detections have
been made, a joint probability is assessed.  These joint probabilities
are used for comparing candidates against each other and generating a
prioritized candidate list for re-observation.

\section*{THE SETI@HOME SKY SURVEY}

SETI@home data comes from the same piggyback receiver that SERENDIP
uses at the Arecibo radio telescope.  Whereas SERENDIP analyzes
this data primarily using a special-purpose spectrum analyzer and
supercomputer located at the telescope, SETI@home records the data, and
then distributes the data through the Internet to hundreds of thousands
of personal computers.  This approach provides a tremendous amount of
computing power, but limits the amount of data that can be handled.
Hence SETI@home covers a relatively narrow frequency range (2.5 MHz),
but searches for a wider range of signal types, and with improved
sensitivity.\cite{Anderson00,Sullivan97}

SETI@home was launched on May 17, 1999.  In its 10 years of operation,
it has attracted over 5 million participants.  SETI@home is one of the
largest supercomputers on our planet, currently averaging 3.5 PFLOP
actual performance.  Users are located in 226 countries, and about 50\%
of the users are from outside of the United States.

Although SETI@home has 1/80 the frequency coverage of SERENDIP V.v,
its sensitivity is roughly ten times better.  The SETI@home search also
covers a much richer variety of signal bandwidths, drift rates, and time
scales than SERENDIP V.v or any other SETI program to date.

Primary data analysis, done using distributed computing, computes power
spectra and searches for ``candidate" signals such as spikes, Gaussians,
and pulses.  Secondary analysis, done on the project's own computers,
rejects RFI and searches for repeated events within the database of
candidate signals.

SETI@home covers a 2.5 MHz bandwidth centered at the 1420 MHz
Hydrogen line from each of the 14 ALFA receivers (7 beams $\times$
2 polarizations).  The 2.5 MHz band is recorded continuously onto SATA
disks with two bits per complex sample.  Disks are mailed to UC Berkeley
for analysis.

SETI@home data disks from the Arecibo telescope are divided into small
``work units" as follows: the 2.5 MHz bandwidth data is first divided
into 256 sub-bands; each work unit consists of 107 seconds of data from
a given 9,765 Hz sub-band.  Work units are then sent over the Internet
to the client programs for the primary data analysis.

Because an extraterrestrial civilization's signal has unknown bandwidth
and time scale, the client software searches for signals at 15 octave
spaced bandwidths ranging from 0.075 Hz to 1220 Hz, and time scales
from 0.8 ms to 13.4 seconds.  The rest frame of the transmitter is
also unknown (it may be on a planet that is rotating and revolving),
so extraterrestrial signals are likely to be drifting in frequency with
respect to the observatory's topocentric reference frame.  Because the
reference frame is unknown, the client software examines about 1200
different Doppler acceleration frames (dubbed ``chirp rates"), ranging
from -100 Hz/sec to +100 Hz/sec.

At each chirp rate, peak searching is done by computing non-overlapping
FFTs and their resulting power spectra.  FFT lengths range from 8 to
131,072 in 15 octave steps.  Peaks greater than 24 times the mean power
are recorded and sent back to the SETI@home server for further analysis.

Besides searching for peaks in the multi-spectral-resolution data,
SETI@home also searches for signals that match the telescope's Gaussian
beam pattern.  Gaussian beam fitting is computed at every frequency and
every chirp rate at spectral resolutions ranging from 0.6 to 1220 Hz
(temporal resolutions from 0.8 ms to 1.7 seconds).  The beam fitting
algorithm attempts to fit a Gaussian curve at each time and frequency in
the multi-resolution spectral data.  Gaussian fits whose power exceeds
the mean noise power by a factor of 3.2 and whose reduced $\chi^2$ of
the Gaussian fit is less than 1.42 are reported to the SETI@home servers.
SETI@home also searches for pulsed signals using a modified Fast Folding
Algorithm\cite{Staelin69} and an algorithm which searches for three
regularly-spaced pulses.  More details of the SETI@home analysis can be
found in Korpela et al.~(2001)\cite{Korpela01} and Korpela (2011)\cite{Korpela11}.

In the near future, we will be adding the ability to detect autocorrelated
signals. Harp et al.\cite{Harp11}~propose that an extraterrestrial
civilization could send a beacon that contains information (has
appreciable bandwidth), yet is easily detectable.  This could be done by
sending a signal and then, after a short delay, starting the broadcast
of a copy of the signal.  A signal of this type can be detected through
autocorrelation, which will show a peak power at the given delay.  Once
the delay is known, the information within the signal can, in principle,
be recovered.  A version of SETI@home containing an autocorrelation
detector, acting on delays up to 13.4 seconds, is being beta tested,
and should be released in late summer of 2011.

To determine signals of interest the data for each sky position which
has recently received new potential signals is examined by our Near-Time
Persistency Checker (NTPCkr).  This program scores candidates based upon
the probability that the assemblage of potential signals seen could be
due to random fluctuations in the noise background.  This score includes
the probability of multiple detections in any reference frame, the
probability of repeated detection in the barycentric frame, the goodness
of fit with the antenna beam pattern, and conicidence with known planets,
nearby stars (from the Hipparcos catalog) or galaxies.	We generate a
ranked list of our best candidates for reobservation\cite{Korpela04}.

Most of the signals found by the client programs turn out to be
terrestrial based radio frequency interference (RFI).  We employ
a substantial number of algorithms to reject the several types
of RFI\cite{Cobb00} from the best signals.  Once RFI rejection
has been performed on a candidate group, it is re-scored by the
NTPCkr.\cite{Korpela11b}

\section*{SETI@HOME--GREEN BANK}

In May of 2011, we utilized the 100m NRAO Green Bank Telescope (GBT) to 
observe 86 of the best planetary system candidates found by the {\em Kepler} 
mission up to that point. We selected distinct {\em Kepler} Input Catalog 
stars having either a {\em Kepler} Object of Interest (KOI) with a calculated 
equilibrium temperature between
230 and 380 K, at least 4 KOIs or a KOI with inferred radius $\lt$ 3.0 Earth
radii and an orbital period $\gt$ 50 days.

Using the GUPPI pulsar 
processor and modified software we were able to record eight 100MHz bands in
dual polarization using 4 bits per complex sample, to provide simultaneous 
coverage between 1.1 and 1.9 GHz.  Following the observations of individual 
candidates for 450 seconds each, we performed a 12 hour survey of the entire 
{\em Kepler} field.   In addition to being analyzed using local 
computing clusters, we will be sending this data out to SETI@home volunteers
for analysis with the SETI@home and Astropulse applications.  
Following some minor modifications to the SETI@home and Astropulse software 
necessary to support the new data parameters, we expect to start 
sending this data out to volunteers starting in the fall of 2011.

\section*{BOINC} 

SETI@home has clearly shown the viability of volunteer based distributed
computing for other scientific problems.  To this end we have developed
an infrastructure dubbed BOINC (Berkeley Open Infrastructure for Network
Computing), which can be used for other applications.\cite{Anderson04}
The availability of this open
source infrastructure has eased the development of other distributed
applications.  Our open source distributed computing infrastructure
engages the public in climate modeling/global warming studies,
(\href{http://climateprediction.net/}{climatePrediction.net}),
HIV, malaria and cancer drug research
(\href{http://www.worldcommunitygrid.org/}{World Community Grid}),
particle physics (\href{http://lhcathome.cern.ch/}{LHC@home}), gravity
waves (\href{http://einstein.phys.uwm.edu/}{Einstein@home}), protein
structure (\href{http://boinc.bakerlab.org/rosetta/}{Rosetta@home}),
and of course SETI@home.  More than 50 distributed computing projects
use this infrastructure.

The BOINC infrastructure has many advantages over the stand alone
SETI@home infrastructure.  Volunteers can sign up for many projects and
divide their computing resources among them.  It is our hope that this
will lead to a larger shared volunteer base.  The BOINC infrastructure
makes it possible for a project to include multiple separate applications
and to distribute updates to applications without requiring complex
user intervention.  The SETI@home BOINC ``project'' can include multiple
``applications'' to analyze data in different ways and to analyze data
from different sources.  Currently we distribute the standard SETI@home
application and the Astropulse application to users who have signed up
for SETI@home.

The BOINC client performs the communications necessary to download
the SETI@home application, to download the data to be analyzed, and to
upload the results.  It also can display screen saver graphics showing
the status of the analysis.
 BOINC and SETI@home are available for MacOS X, Windows and
Linux in both 32 and 64 bit versions.  Participants using other
operating systems can download the source code, and can compile
their own versions of BOINC and SETI@home.  Participants can download
the client software at: 
\href{http://setiathome.berkeley.edu/}{\em http://setiathome.berkeley.edu/}

\section*{FLY'S EYE}

Our now completed "Fly's Eye" experiment searched for bright radio
transient pulses at the Allen Telescope Array (ATA).  The Allen
Telescope Array has several advantages over other telescopes worldwide
for performing transient searches, particularly when the search is for
bright pulses. The ATA has 42 independently-steerable dishes, each 6m in
diameter.  The beam size for individual ATA dishes is considerably larger
than that for most other telescopes, such as VLA, NRAO Green Bank, Parkes,
Arecibo, Westerbork and Effelsberg. By pointing each dish in a different
direction the ATA can instantaneously observe a far larger portion of the
sky than is possible with other telescopes. Conversely, when using the
ATA dishes independently, the sensitivity of the ATA is far lower than
that of other telescopes.  The Fly's Eye instrument was built to
search for bright radio pulses of submillisecond duration at the ATA. The
instrument consisted of 44 independent spectrometers each processing
a bandwidth of 210MHz, and producing 128-channel power spectrum at a
rate of 1600Hz. Therefore each spectrum represents time domain data of
length 0.625 ms, and hence pulses as short as 0.625 ms can be resolved

Because interstellar pulses are dispersed by ionized interstellar gas,
a pulse search involves correcting for this dispersion.  The analysis
required for the Fly's Eye experiment is, in principle, fairly simple. 
We wish to search over a wide range of dispersion measures to find
large individual pulses. Specifically our processing requires that all
the data be dedispersed with dispersion measures ranging from 50 \dm to
2000 \dm. At each dispersion measure the data needs to be searched for
'bright' pulses.  The processing chain is in practice significantly
more complicated than this description suggests. Processing is performed
on compute clusters, with input data formatted, divided and assigned
to worker nodes for processing. In the worker node flow, the data is
equalized, RFI rejection is performed, and finally a pulse search is
performed through the range of dispersion measures. The results are
written to a database where they can be subsequently queried. The key
feature of the results is a table that lists, in order of decreasing
significance, the pulses that were found and their dispersion measures.

We have been able to observe 150 deg$^2$ over 450 hours.
We have successfully detected three pulsars (B0329+54, B0355+54,
B0950+08) and six giant pulses from the Crab pulsar. We have not
detected any other convincing bursts of astronomical origin in our
survey data implying a limiting rate of less than 2 sky$^{-1}$
hour$^{-1}$ for 10 ms duration pulses having average apparent 
flux densities greater than 44 Jy.  More information about "Fly's Eye" 
is available in Siemion et al.~(2010,2011)\cite{Siemion10,Siemion11}

\section*{ASTROPULSE}

Radio SETI searches to date have concentrated on narrow-band signals as
opposed to wide-band signals such as pulses.  The Astropulse project
was the first SETI search for $\mu$s radio pulses.  Astropulse detects
pulse widths ranging from 1 $\mu$s to 1 ms.  Such pulses might come
from extraterrestrial civilizations, 
evaporating black holes,\cite{Hawking74,Rees77}
gamma ray bursts, 
certain supernovae, 
or pulsars.\cite{Cognard96}  The Astropulse program mines
the SETI@home data archive for serendipitous detections of such events.

One of the unique features of this search is that it is the first pulse
search to use coherent dedispersion in a ``blind'' fashion - we have no
previous knowledge of a specific dispersion measure (DM) to examine.
The reason this search has never been attempted before is due to the
enormous computing power required.

The computing problem is eminently parallel in nature.	Similar to
SETI@home, Astropulse uses volunteers and their personal computers
to carry out the computation.  Astropulse uses the general purpose
distributed computing system we have developed (BOINC).

Thus far, Astropulse has examined several years of SETI@home data,
resulting in more than 125 million potential detections.  We are
currently working on RFI rejection techniques to clean the data set of
RFI, primarily due to aviation radars.	Most RFI sources appear to have a
stable dispersion measure. If we detect a pulse with the same dispersion
measure, but from a different part of the sky minutes to hours later,
it is likely to be RFI. In practice we compute the likelihood of such
coincidence happening and if that likelihood is below a threshold we
conclude that it is RFI.  We use two more steps to further select the best
pulses. First we check to see that the pulse is truly broadband. Some
pulses can be due to chance coincidence of peaks at two frequencies
with a time offset. We eliminate these by checking the distribution of
power vs frequency and using a $\chi^2$ test from a fit with a flat spectrum
to determine whether it is a broad band pulse.	Finally, because we have
two polarizations being analyzed independently we can, at the cost of
some sensitivity, require that the pulse be detected in both polarization.

Using this procedure on the first 12 million pulses we found, 330 pulses
from 114 sky locations passed the test.  Two of these locations correspond
with known pulsars, leaving 112 that do not.  These candidates are
concentrated near the Galactic plane, much more than our observing time
has been.  A statistical analysis\cite{vonKorff10} indicates a greater
than 4$\sigma$ probability that this distribution is not due to chance.
We reobserved most of these candidate locations in July 2011.  However,
given the volume of data, our analysis of these reobservations is just
beginning.

More details of Astropulse
can be found in von Korff (2010)\cite{vonKorff10} and von Korff et al (2011).\cite{vonKorff11}

\begin{figure}[tb] 
\begin{center} 
\ifpdf
\epsfig{file=lab_map.pdf,width=6.5in}
\else
\epsfig{file=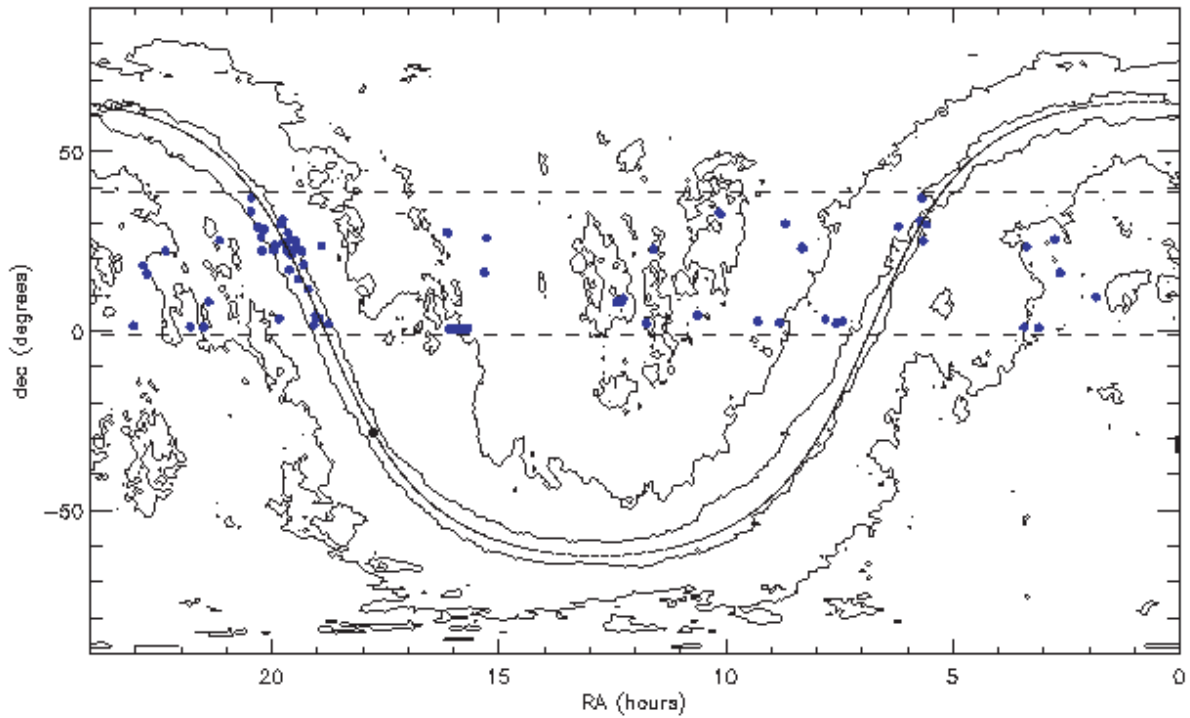,width=6.5in}
\fi
\end{center} 
\caption{Distribution of Astropulse candidates on
the sky.  The dots are the potential sources. Contours are LAB HI
survey\cite{LabSurvey} logarithmic contours.  The dashed horizontal
lines represent the limits of Arecibo's viewing range.} 
\end{figure}

\acknowledgements

This work has been supported by the Planetary Society, the SETI
Institute, the University of California, Sun Microsystems and donations
from individuals around the planet.  Key hardware was donated by Network
Appliance, Xilinx, Fujifilm Computers, Toshiba, Quantum, Hewlett Packard,
and Intel Corp.  We receive excellent support from the staff of the
Arecibo Observatory, a part of that National Astronomy and Ionosphere
Center, which is operated by Cornell University under a cooperative
agreement with the National Science Foundation.  We would also like
to thank the Allen Telescope Array, a facility of the SETI Institute.
The National Radio Astronomy Observatory in Green Bank, WV is a 
facility of the National Science Foundation operated under 
cooperative agreement by Associated Universities, Inc. 
This work has been supported in part by NSF Grants AST-0808175 and
AST-0307956, and NASA Grant NNX09AN69.

\bibliography{korpela} \bibliographystyle{spiebib}

\end{document}